# Oscillator Based on Lumped Double Ladder Circuit with Band Edge Degeneracy

Dmitry Oshmarin, Farshad Yazdi, Mohamed A. K. Othman, Jeff Sloan, Mohammad Radfar, Michael M. Green, and Filippo Capolino

*Abstract*—A new oscillator design based on a periodic, double ladder resonant degenerate circuit is proposed. The circuit exhibits a degenerate band edge (DBE) in the dispersion diagram of its phase-frequency eigenstates, and possesses unique resonance features associated with a high loaded *Q*-factor resonance, compared to a single ladder or a conventional LC tank circuit. This oscillator is shown to have an oscillation threshold that is half that of a single LC ladder circuit having the same total quality factor, and thus is more robust than an LC oscillator in the presence of losses. Moreover, the double ladder oscillators have a unique mode selection scheme that leads to stable single-frequency oscillations even when the load is varied. It is also shown that the output amplitude of the double-ladder oscillator is much less sensitive to the output loading compared to single-ladder oscillators. We show the analysis and design of such oscillators that potentially lead to enhancing the efficiency of RF components and sources.

## I. Introduction

Oscillators are essential components of any radio frequency (RF) system. Typically, an RF oscillator operates via a positive feedback mechanism utilizing a gain device with a selective resonance circuit that generates a single tone used as the carrier frequency. Van der Pol topologies are among the most conventional oscillators utilized in RFs thanks to their simplicity of design and ease of integration [1]. Indeed, most voltage-controlled oscillators (VCOs) are designed based on an LC-tank circuit [2]. The negative conductance required for positive feedback can be obtained by simple circuit structures such as a cross-coupled transistor pair [2]. A negative conductance can be also obtained from other circuit topologies such as Pierce, Colpitts, and Gunn diode waveguide oscillators [3]–[5]. While used widely, all designs based on an LC-tank circuit have some important limitations; in particular, their performance largely depends on loading conditions. They often require one or more power-hungry buffer stages to terminate the signal to a low output impedance (often 50 Ω), which can be undesirable for low-power applications.

Pursuing better performing RF and microwave sources is an important research avenue where novel principles of RF generation are currently being investigated [6]–[10]. Other designs may feature distributed [8], [11], coupled [12] or multi-mode [6] oscillators. The focus of this paper is on a particular class of oscillators whose architecture features a repeating cascade of unit cells, each consisting of reactive components. One of the advantages of such a circuit is that its criteria for oscillation are more relaxed and its oscillation frequency is independent of loading, which will be discussed in Section II. In addition, the proposed circuit potentially offers a significantly more power-efficient way to terminate the output signal with low-impedance loads, as it does not need output buffer stages for the load termination. In Section II we provide a brief description of the properties of single- and double- ladder circuits. In Section III we explore the resonance characteristics including the effect of losses of the double ladders. In Section IV we analyze the threshold conditions for oscillation by a negative differential resistance that can be realized by a cross-coupled differential CMOS transistor pair. In Section V we investigate time-domain behavior of the oscillators that includes the active device nonlinearity. In Sections II, III, and IV we assume that the circuits are operating in the sinusoidal steady-state so that all voltages and currents are represented by phasors. All the calculations in the paper, except for those in Section IV, are carried out using a formalism based on the four-dimensional state vector $\Psi_n = [V_1(n) \quad V_2(n) \quad I_1(n) \quad I_2(n)]^T$, (pertaining to Fig. 1(c)), and the corresponding 4×4 transfer matrix of a unit cell **T** as detailed in [13]. Time-domain simulations in Section IV are carried out using Keysight Advanced Design System (ADS).

## II. Single- and Double- Ladder Circuits

### A. The Single-Ladder Circuit

We consider a periodic resonant circuit made up of LC ladder cells connected in tandem. A simple example of such a cell, comprising a series inductor and a shunt capacitor, is shown in Fig. 1(a). Finite-length implementations of periodic circuits in

This material is based upon work supported by the Air Force Office of Scientific Research under award number FA9550-15-1-0280 and under the Multidisciplinary University Research Initiative award number FA9550-12-1-0489 administered through the University of New Mexico.

The Authors are with the Department of Electrical Engineering and Computer Science, University of California, Irvine, CA 92697 USA. (e-mail: doshmari@uci.edu, fyazdi@uci.edu, mothman@uci.edu, mradfar@uci.edu, mgreen@uci.edu, f.capolino@uci.edu)

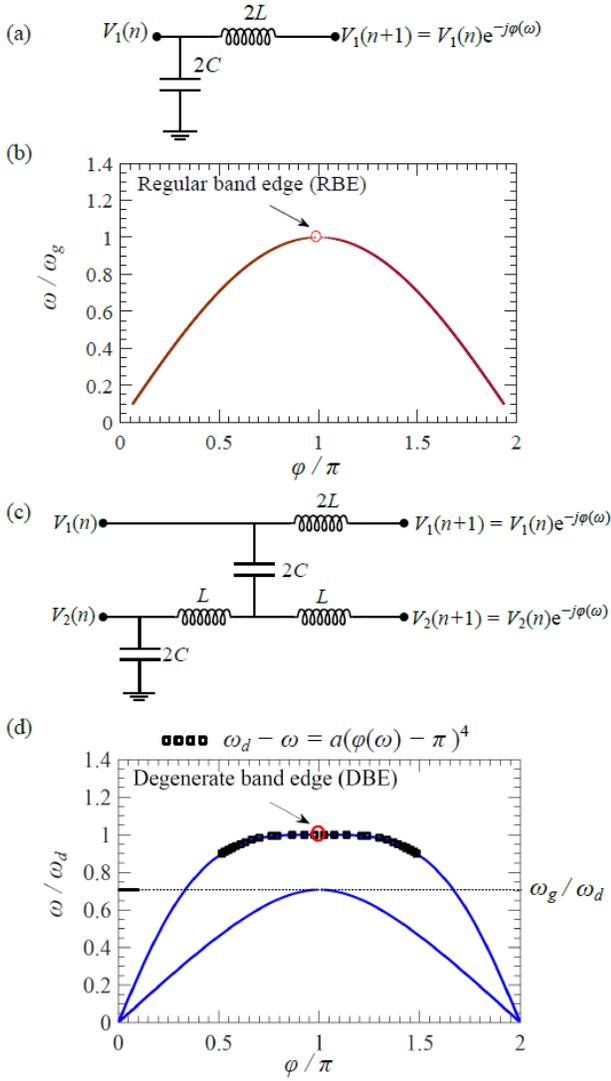

Fig. 1. (a) Unit cell of a periodic single-ladder lumped circuit; (b) dispersion diagram of the periodic single-ladder circuit that develops an RBE at an angular frequency $\omega_g$; (c) unit cell of a periodic double-ladder lumped circuit; (d) dispersion diagram of the periodic double-ladder circuit that develops a DBE at an angular frequency $\omega_d$. Also, the asymptotic dispersion relation typical of DBE, $\omega - \omega_d = a[\varphi - \pi]^4$, is plotted as square symbols.

practice have many applications including filters, pulse shaping networks, and delay lines [14]–[16].

Let us now suppose that the Fig. 1(a) ladder is of infinite length, is excited by a sinusoidal source with angular frequency $\omega$, and is operating in the steady state. For this mode of operation all of the currents and voltages can be expressed as phasors; in particular, the phasor ratio $V_1(n+1)/V_1(n)$ can be determined by simply finding the eigenvalues and eigenvectors of the transmission matrix describing the unit cell (note that these eigenvalues are not the same as the circuit's natural frequencies; see Ch. 8 in [17]). Moreover, it can be shown that for sufficiently low angular frequency $\omega$, $V_1(n+1)/V_1(n)$ has unity magnitude; i.e., all voltages and currents have the same amplitude, differing only by a fixed phase shift $j(W)$. Therefore, we define the eigenstates for a periodic circuit as the possible solutions of the eigenvalue problem describing the evolution of the voltages and currents from one cell to the next in the periodic circuit following the Bloch-Floquet theory (see Ch. 8 in [17], and also [18]). In general, each eigenstate is characterized by voltages and currents that from cell to cell vary as $\exp(-j\gamma)$, where $\gamma = \varphi - j\alpha$ is the complex phase shift from one unit cell to the next.

The characteristic between the applied frequency and the phase shift between cells in an infinitely long periodic ladder is known as the *dispersion relation*. For the Fig. 1(a) ladder, it can be shown that the eigenvalues of the transfer matrix [13], [17] are given by

$$e^{-j\gamma} = 1 - 2\left(\frac{\omega}{\omega_g}\right)^2 \pm 2\frac{\omega}{\omega_g}\sqrt{\left(\frac{\omega}{\omega_g}\right)^2 - 1} \quad (1)$$

where $\omega_g = \sqrt{1/(LC)}$. For $\omega = \omega_g$, these eigenvalues have unity magnitude at which $\gamma = \varphi = \pi$ and $\alpha = 0$. The corresponding dispersion relation is shown in the curve labeled regular band edge (RBE) in Fig. 1(b) at $\omega_g = \sqrt{1/(LC)}$. (In this curve the range of principal values of the inverse tangent function used to obtain $\varphi(\omega)$ is chosen to be from 0 to $2\pi$, and only the dispersion of eigenstates for which $\alpha = 0$ are plotted as conventionally done [17]). The angular frequency $\omega_g$, known as the *band edge*, defines the passband for the periodic circuit. At frequencies higher than $\omega_g$ the eigenvalues will no longer have unity magnitude, and thus $V_1(n+1)/V_1(n)$ will have both an attenuation factor and a phase shift. This range of frequencies corresponds to the stopband, and the condition is known as an evanescent state. The dispersion of these states of phase near $\omega = \omega_g$ behaves as $(\omega_g - \omega) \propto (\gamma - \pi)^2$.

Note that in Fig. 1(b) at $\omega = \omega_g$ the two phase shifts corresponding to the eigenvalues in (1) coalesce into a single value. This phenomenon is well-known in periodic structures that naturally exhibit an electromagnetic band gap [17]. In general, periodic structures composed of transmission lines or waveguides have been shown to demonstrate unique properties associated with "slow-light" properties near the band edge [18]–[21]. Such properties are associated with very high group delay near the band edge and consequently lead to enhancing the quality factor of resonators. Several applications has been investigated in lasers and high-power electron beam devices based on the band-edge operation [19], [22]–[24].

*B. The Double-Ladder Circuit*

The proposed oscillator in this paper is based on a periodic double-ladder circuit whose unit cell with four ports is shown in Fig. 1(c). Since each unit cell is a grounded four-port network, this circuit supports four eigenstates rather than two, as was the case for the unit cell in Fig. 1(a) [5], [25]. Under some particular choice of the circuit elements, at a certain frequency these four eigenstates coalesce at $\varphi = \pi$, $\alpha = 0$ resulting in the so-called degenerate band edge (DBE) condition [21], [26]–[28]. The four sets of voltage and currents associated with these eigenstates are no longer independent at this degeneracy point [13]. A DBE can only be found in double ladders since it represents the degeneracy of four eigenstates [13]. Near the DBE condition, the complex phase-frequency dispersion relation of these states is characterized by $(\omega_d - \omega) = a(\gamma - \pi)^4$ where $a$ is a geometry-dependent fitting constant, and was found analytically in [13].

For the circuit in Fig. 1(c), the DBE angular frequency is given by $\omega_d = 1/\sqrt{LC}$ while the characteristic impedance is $Z_c = \sqrt{L/C}$ [13]. The theory of lumped circuits with DBE has been developed in [13], where different double-ladder circuit configurations composed of cascaded identical unit cells are studied. The normalized dispersion diagram for the four-port periodic circuit with the proposed unit cell is depicted in Fig. 1(d). At $\omega = \omega_g = \omega_d/\sqrt{2}$ two of the eigenstates exhibit an RBE, similar to the dispersion of a single LC ladder. However, at $\omega = \omega_d$ a DBE where all four states coalesce can be observed. For the circuit in Fig. 1(c) a value of $a = 120$ s$^{-1}$ was used to fit the curve in Fig.1(d). To provide a practical implementation at RF frequencies values for capacitors and inductors are chosen from commercially available lumped elements whose $Q_e$ factor can exceed 500. Namely, in the following the ladder circuit is composed of inductors and capacitors whose values are $L = 45$ nH and $C = 56$ pF, respectively. As a result, the circuit has a DBE frequency $f_d = 1/(2\pi\sqrt{LC}) \cong 100.26$ MHz and $Z_c = 28.3$ Ω.

### III. RESONANCES OF PASSIVE DOUBLE LADDER CIRCUIT

In this section, we consider double-ladder circuits as shown in Fig. 2(a) made by cascading a finite number $N$ of unit cells shown in Fig. 1(c), and analyze their resonance characteristics related to the DBE. In particular, we first investigate important characteristics of passive double-ladder circuits, for which the effects of element losses on the transfer functions, loaded quality factor, and driving point admittance are explored in detail. These particular features allow for an unconventional way to construct oscillators. We also compare the double-ladder and single-ladder oscillators, while highlighting the advantages of the former, and demonstrate that single-ladder oscillators can operate based on multiple resonant modes and thus can generate multiple frequencies, while double-ladder oscillators can only oscillate at a single frequency.

Another undesirable property of the single-ladder configuration is mode jumping, in which the frequency of oscillation changes with the load variation, as reported in the literature [6]. However, here we demonstrate that double ladder oscillators are not prone to load variations, thereby exhibiting a more stable oscillation frequency and lower threshold for oscillation as compared to a single-ladder implementation.

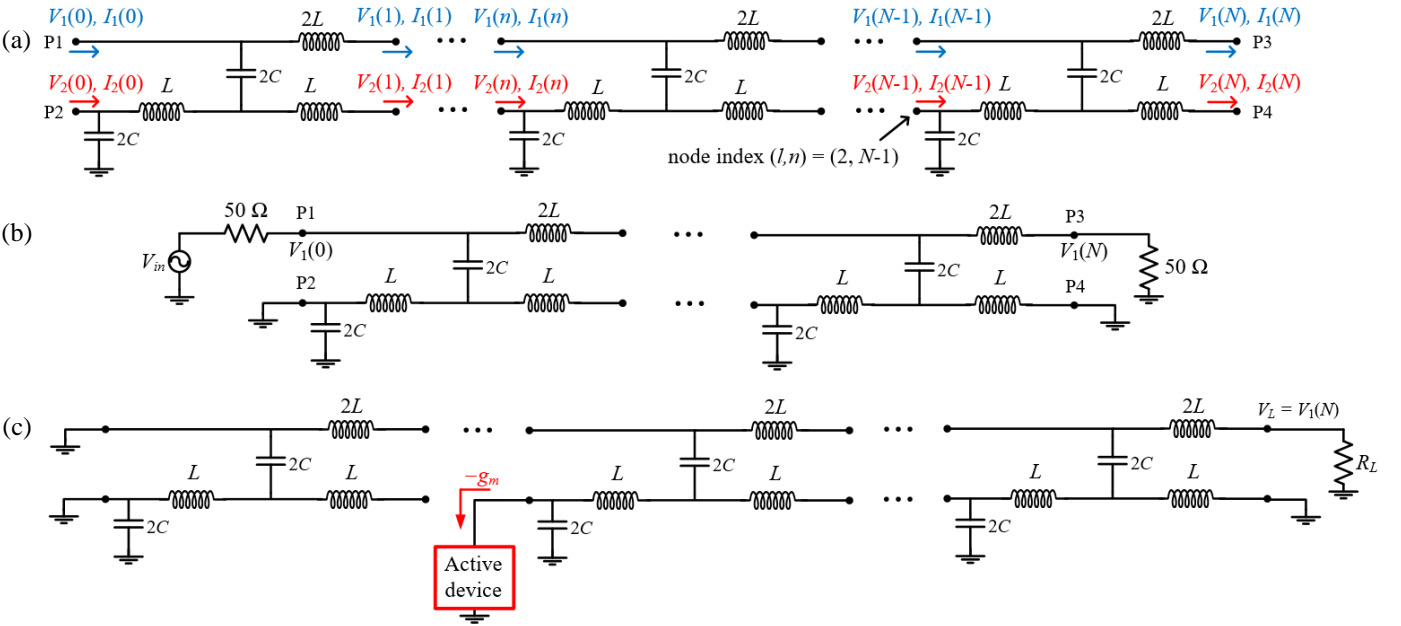

Fig. 2. (a) Double LC ladder periodic circuit made of $N$ unit cells operated near the DBE. P1 through P4 represent the four ports of the $N$-cell circuit; (b) circuit with excitation voltage $V_{in}$ and source resistance of 50 Ω, and 50 Ω load at P3; (c) double-ladder oscillator with terminations and active device configuration (other configurations may have an active device in each unit cell).

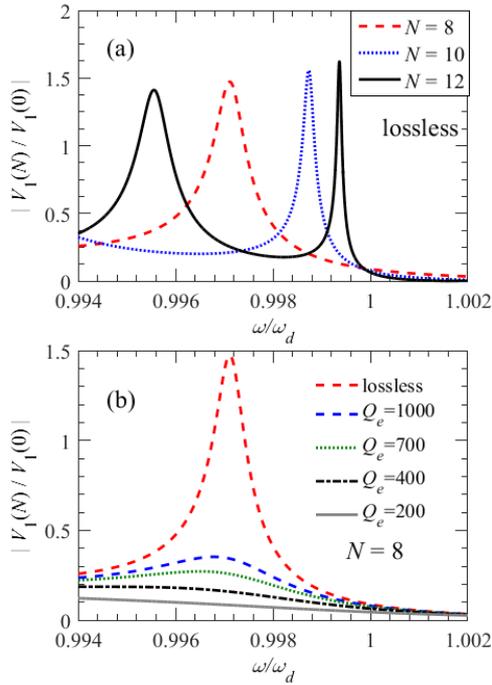

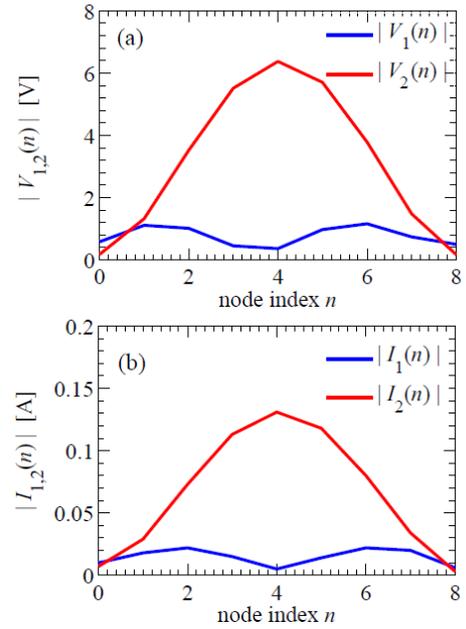

Fig. 3. The voltage transfer function between the upper output P3 and upper input P1 nodes of the circuit with terminations as shown in Fig. 2(b) for (a) different number of unit cells and no loss in the elements and (b) different quality factors for elements in an 8 unit cells resonator. The important resonance is the one close to $\omega_d$, denoted as $\omega_{r,d}$.

Fig. 4. (a) Voltage distribution corresponding to $Q_e = 400$ in the upper ladder nodes and the lower ladder nodes of the circuit composed of 8 unit cells at the DBE resonance occurring at $\omega_{r,d}$; (b) current distribution in the upper ladder nodes and the lower nodes of the circuit composed of 8 unit cells.

As shown in Fig. 2(a), every unit cell has four terminal nodes, each of which is identified by two indices: $l \in \{1,2\}$ denoting the upper or lower ladder, respectively, and $n \in \{1,2,\ldots,N\}$ denoting the node location along the double ladder. As such, each cell's terminal nodes in the Fig. 2(a) ladder will be referred to using the notation $(l,n)$. When a double-ladder circuit composed of $N$ cells is terminated at both ends by resistive loads, there will be several complex-valued natural frequencies, corresponding to resonance modes. It can be shown that the imaginary part of the natural frequencies $\omega_{r,m}$ near the DBE angular frequency $\omega_d$ can be approximated by $\omega_d - \omega_{r,m} \approx (m/N)^4$ where $m$ is a positive integer designating the resonance mode, observable from the peaks in Fig. 3 (similar to other periodic structures with DBE in [26], [29], [30]). We focus here on the closest resonance to the DBE corresponding to $m = 1$ with imaginary part $\omega_{r,1}$, since it is associated with an excitation of all the eigenstates supported by the periodic ladder (i.e., the four eigenvalues are nearly identical near the DBE as illustrated in Fig. 1(d)). To be consistent with [13] we will refer to this angular frequency as $\omega_{r,d}$ throughout the rest of this paper.

Characteristics of the DBE resonator have been discussed thoroughly in [16]−[21] and in particular double-ladder lossless circuit properties have been shown in depth in [13]. Here, instead, we investigate in detail the effect of losses on the performance of the DBE resonator and how they are related to threshold criteria for oscillation and the performance on DBE-based oscillators in general.

### A. Transfer function

As mentioned in the previous Section, the DBE resonance mode is associated with an excitation of all four eigenstates [26], [28], [29]. In this Section we assume, as shown in Fig. 2(b), that the double ladder is terminated with 50Ω at port 3, and with a voltage source $V_{in}$ in series with 50Ω at port 1. The lower ladder is shorted to ground at ports 2 and 4. We first calculate the voltage transfer function $|V_1(N)/V_1(0)|$ of this circuit for frequencies near the DBE, for $N = 8$, 10, and 12; the magnitudes of these transfer functions are shown in Fig. 3(a). As shown in [13], the DBE-related resonance peak at $\omega_{r,d}$ in the double-ladder circuit exhibits the narrowest transmission peak compared to the other resonances, and its quality factor has been shown to scale as $N^5$ [13]. As noticed previously [28], near the DBE the group delay is very large as can be observed by the flat region of the dispersion diagram in Fig. 1(d), which corresponds to a high quality factor even when the circuit is terminated by its characteristic impedance. (This impedance is the result of the chosen L and C for oscillation frequency at 100 MHz, as described in Sec. II.B.). Furthermore, as can be seen from Fig. 3(a), by increasing the number of unit cells the DBE resonance angular frequency $\omega_{r,d}$ approaches the DBE angular frequency $\omega_d$. However, when losses are introduced into each L and C in the circuit the $\omega_{r,d}$ resonance loaded quality factor significantly declines, compared to other resonances of the circuit. In Fig. 3(b) the transfer function of the double ladder is depicted only for the resonance closest to the DBE for $N = 8$. We assume that all elements have the same quality factor $Q_e$ for simplicity. Increasing losses beyond a certain limit deteriorates the DBE resonance, and the transfer function's peak can completely vanish for sufficiently low $Q_e$, as for the case with $Q_e = 200$ in Fig. 3(b).

Fig. 4 shows how voltage and current magnitudes at the angular frequency $\omega_{r,d}$ vary throughout the circuit for $Q_e = 400$. These voltages and currents are evaluated at the $n$th node, $n = 1,2,\ldots,8$, in the finite double-ladder circuit in Fig. 2(b). In Fig. 4(a) the voltage distributions on the lower transmission line (red) and the upper transmission line (blue), are depicted; likewise, in Fig. 4(b) the current distributions for the upper and lower lines are plotted. It is interesting to note that both the voltage and current reach their peak magnitudes in the middle of the lower ladder; these magnitudes are approximately six times larger than those of the upper ladder, even when losses are present. On one hand, the reason for excitation of such voltage and current in the resonator is due to the excitation of all of the eigenstates of the periodic double ladder near the DBE condition, which is a general property of DBE resonators [26], [28]–[30]. On the other hand, the eigenstates of such particular periodic ladders have a voltage distribution that is mostly confined to the lower ladder, in the sense that the upper ladder nodes are essentially RF grounds; this is also true of the middle node in the lower ladder (node with $l=2$, $n=4$) [13]. The same behavior can be seen for the current distribution as well. Therefore, most of the energy stored in the resonator is confined in the lower ladder's components.

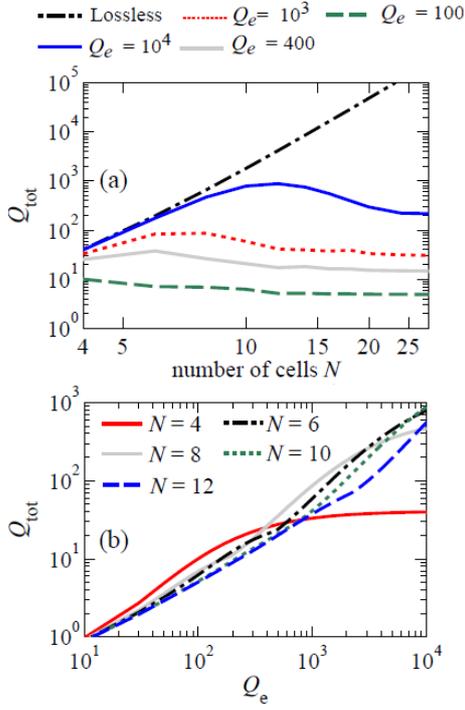

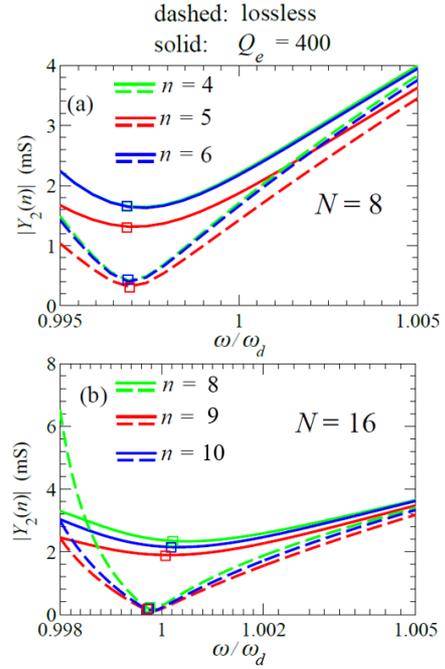

Fig. 5. (a) Loaded quality factor, $Q_{tot}$, versus number of unit cells for different element quality, as well as the lossless theoretical limit. (b) Loaded quality factor, $Q_{tot}$, versus element quality factor for different number of unit cells.

Fig. 6. Magnitude of admittance near the DBE throughout the different nodes of the double ladder circuit for (a) $N = 8$ unit cells and (b) $N = 16$ unit cells. Square symbols denote the resonance frequency $\omega_{r,d}$ at which $\text{Im}(Y_2(n)) = 0$.

### B. Total quality factor

To provide a comprehensive analysis of the performance of the lossy resonator, we calculate the loaded quality factor $Q_{tot}$ of the double ladder; i.e., the quality factor of the circuit including the resistive port terminations as well as the losses in the L and C components. This loaded quality factor associated to the resonance $\omega_{r,d}$ is defined as [17]

$$Q_{tot} = \omega_{r,d} \frac{W_e + W_m}{P_l} \qquad (2)$$

where $W_e$ and $W_m$ are the total time-average energy stored in the circuits in the capacitors and inductors, respectively, and $P_l$ is the time-average power dissipated in the resistive terminations and in the components' loss. In [13], the loaded quality factor of a lossless double ladder is thoroughly analyzed, and here instead we focus on the effect of component losses on $Q_{tot}$. For the lossless case $Q_{tot}$ is proportional to $N^5$ for large $N$ as shown in [13] for this circuit and for other structures with DBE [28], [33], [34]. However, for the case when the reactive components are lossy, the loaded quality factor does not grow indefinitely as $N^5$ but it is limited by the loss in the elements. Therefore, as seen in Fig. 5(a), the quality factor for the lossy cases grows exponentially only for small $N$, then declines and saturates for larger $N$. For $Q_e = 400$, $Q_{tot}$ shows little variation with $N$ since it is already very low. The effect of element losses limiting the loaded quality factor is further explored in Fig. 5(b), where we show $Q_{tot}$ versus $Q_e$ for different values of $N$. In this plot $Q_{tot}$ increases linearly in and then tends to saturate due to the 50 Ω port terminations. Note that, as discussed in [13], using a very high or very low impedance load may be attractive to enhance the $Q_{tot}$ of the resonator, but would reduce the amount of power available at the terminations. Therefore, the results here are calculated for 50 Ω terminations, as is commonly used in RF circuits.

### C. Driving point admittance $Y_{in}$

The most important characteristic for estimating the oscillation criteria is the driving-point impedance at the location where an active device will be connected in order to start the oscillation [2]. Because of the very large resonance voltage and current in the lower ladder relative to the DBE seen in Fig. 4, driving the lower ladder, especially near the middle of the resonator at the node denoted by $(l,n)=(2, N/2)$ with a negative conductance, would have the greatest impact in compensating the effect of losses in the circuit to achieve oscillation. Note that in other circuit configurations, the same analysis is necessary to predict the location of the peak voltage and consequently the driving point. To provide a meaningful assessment of the oscillation threshold, we consider a single-ended loading scheme in which port 3 is terminated by 50 Ω load, while ports 1, 2 and 4 are terminated by a short circuit as shown in Fig. 2(c).

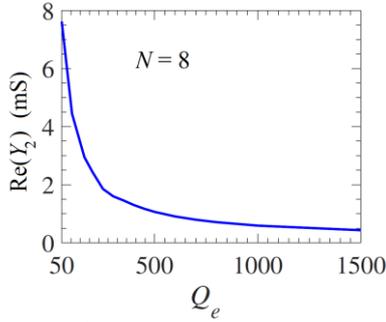

Fig. 7. Magnitude of the driving point admittance at the resonance angular frequency $\omega_{r,d}$ versus quality factor of elements for $N = 8$ cell double ladder resonator.

In Fig. 6 we show the magnitude of the driving-point admittance versus normalized frequency at the three middle nodes of the circuit for $N=8$ and $N=16$ unit cells (i.e., nodes $n = 4,5,6$ and $n = 8,9,10$, respectively, with $l=2$). Both lossless and lossy elements with $Q_e = 400$, are considered. From the results shown in Fig. 6 it can be seen that the lower ladder's node in the 5th and 9th unit cell, corresponding to $N=8$ and $N=16$ respectively, exhibit the lowest input admittances; thus they constitute the appropriate driving points for both configurations. It can also be seen from Fig. 6 that as the number of cells increases, the trend for the input admittance is different for the lossy and lossless circuit.

When $N$ increases from 8 to 16, the admittance decreases by 50% for the lossless case (dashed lines in Fig. 6), while it increases by 40% for the case with $Q_e = 400$ (solid lines). (Note that the minimum value of the admittance's magnitude does not exactly correspond to the resonance condition at which Im($Y_{in}$)=0); the resonance angular frequency $\omega_{r,d}$ is indicated in Fig. 6 with a square symbol.

To realize an oscillator, we use the circuit shown in Fig. 2(c), which shows all of the port terminations and the placement of the active device, and plot the magnitude and the real part of the input admittance at resonance for an 8-cell double-ladder circuit as a function of the loaded quality factor of elements, shown in Fig. 8. It can be observed that the admittance rapidly decreases as $Q_e$ increases, and saturates after some value (for $Q_e < 2000$, we have Re($Y_2$)=0.2 mS). Thus, selecting elements with very high $Q_e$ is not necessary when constructing an oscillator since the applied negative resistance value required for oscillation will be only slightly higher in magnitude. (To compare these characteristics with those of a single LC ladder filters, we refer the reader to [6], [15], [16] for filtering characteristics of a single ladder, as well as for the scaling of the quality factor of regular band edge resonators.)

It is very important to point out the DBE oscillator could operate in two different ways. One would be by applying an active device to each unit cell. This could be done by using a set of differential cross-coupled pairs, for instance. Thus working near the DBE would not only provide for good conditions for the driving point admittance, but would also provide nearly 180º phase shift from one cell to the next such that a cross-coupled pair between two cells would guarantee fully-differential operation. The other way would be to connect just one single-ended active device (such as those in [35], [36], for example) which we investigate next.

### IV. ACTIVE DOUBLE LADDER CIRCUIT

In this section, we investigate the oscillation condition of the double-ladder circuit and compare some of the important characteristics of oscillations of the proposed double-ladder oscillator to a single-ladder-based oscillator. For purposes of the comparison, we will use the same number of cells in both structures. The proposed oscillator is composed of a double ladder terminated by a single-ended resistive load in the upper ladder end as seen in Fig. 2(c) (as described in Section II.C), while the active device (a single-ended negative resistance) is attached to the driving point. As in most LC-based oscillators, the conditions for oscillation are formulated using the Barkhausen criteria for the feedback system [2], [37]. In our case with the double-ladder circuit, the resulting oscillation frequency is not exactly the same as $f_d$ or $f_{r,d}$ due to the possible nonlinearities of the active device, but when increasing the number of unit cells, those two frequencies almost coincide as well as the oscillation frequency.

An active device used to induce oscillations can be characterized by its operating *I-V* curves. A negative resistance can be practically implemented by CMOS transistors or diodes and an example for a third-order I-*V* characteristic is shown in Fig. 8, which utilizes the following equation:

$$I = -g_m V + \alpha V^3, \quad (3)$$

where $-g_m$ is the slope of the *I-V* curve in the negative resistance region, and $\alpha$ is the third-order nonlinearity constant that models the saturation characteristic of the device. (It is this saturation characteristic that determines the steady-state oscillation amplitude.) To realize a constant dc voltage-biased active device we choose the turning point $V_b$ of the *I-V* characteristics to be constant under different biasing levels. In particular, we set $\alpha = g_m / (3V_b^2)$, as shown in Fig. 8. We also assume that the capacitances in the ladder are much larger than any parasitic capacitance associated with the negative resistance device. Calculations are carried out using ADS transient solver.

#### A. Starting oscillation conditions

Conditions for instability that lead to oscillation are found using the pole-zero analysis of the linearized circuit [2]. For simplicity, we consider the complex poles of the voltage transfer function of the double-ladder circuit. In general, the transfer function between the Port 1 and Port 3 voltages, i.e., $V_1(N) / V_1(0)$, can be written as:

$$\frac{V_1(N)}{V_1(0)} = K \frac{(s-z_1)(s-z_2)\cdots(s-z_M)}{(s-p_1)(s-p_2)\cdots(s-p_L)}, \quad (4)$$

where $z_i$, with $i = 1,2,\ldots,M$ and $p_k$, with $k = 1,2,\ldots,L$ are the complex zeros and poles, respectively of the system.

These poles and zeros, together with the gain constant *K,*

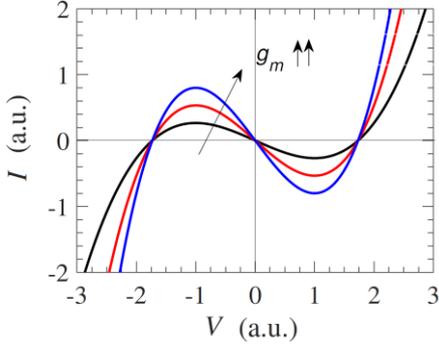

Fig. 8. I–V Characteristic for the active device $I = -g_m V + \alpha V^3$ in which $\alpha = g_m/(3V_b^2)$, that exhibits small-signal negative resistance.

completely characterize the small-signal behavior of the circuit. In our case, we are most interested in the pair of complex-conjugate poles that is the closest to the DBE frequency. The pole configurations for both a double-ladder and a single-ladder circuit are shown in Fig. 9, for the load configuration discussed above. By increasing the loss in the elements of the circuit, the pair of poles closet to the DBE (shown in the inset of Fig. 9) will recede from the imaginary axis into the left-half plane, which can result in a decrease of the peak amplitude of the voltage transfer characteristic. Applying a negative transconductance across the appropriate nodes of the circuit will pull the poles associated with the DBE toward the right-half plane. Since the $(l,n)=(2, N/2)$ node of a circuit has the lowest driving-point admittance, we attach the active device to that node and investigate the minimum transconductance $g_{m,\min}$ needed to start the oscillation as follows.

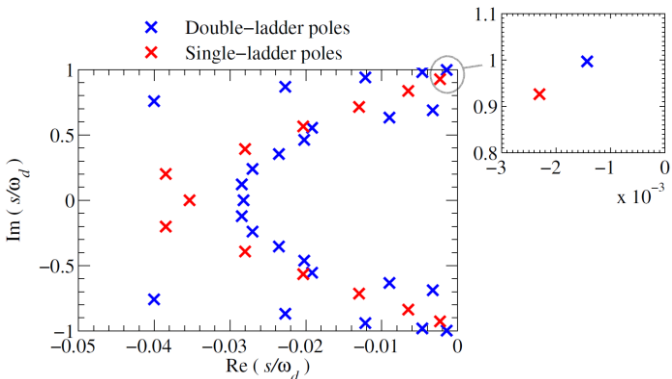

Fig. 9. Pole-zero configuration of the transfer function near imaginary axis for single-ladder and double-ladder circuits where for $N=8$ and lossless elements.

Fig. 10 shows how adding negative resistance to the circuit affects the real part of the poles near the DBE. For each circuit the value of $g_{m,\min}$ corresponds to when this real part is zero. Increasing $g_m$ beyond this threshold allows oscillations to start up. In Fig. 10 the trajectory of the poles near the DBE is shown as a function of $g_m$. We plot the real part of the conjugate poles near the DBE frequency for two cases: In the first (Fig. 10(a)), the negative $g_m$ is placed in all unit cells of the periodic circuit; in the second (Fig. 10(b)), it is applied only to the 5th unit cell. It can be observed that by increasing the value of $g_m$ the real part of the poles goes from negative to positive, which indicates that the pole has crossed the imaginary axis in the s-plane to the right half-plane (RHP).

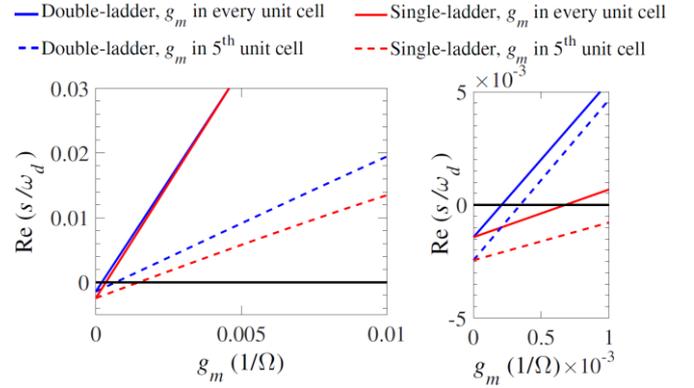

Fig. 10. Real part of the poles near DBE frequency varying as a function of $g_m$ for single-ladder and double-ladder circuits where $N=8$ unit cells. The right panel plot is zoom-in showing where poles transitions into the unstable region.

Moreover, when we compare the double ladder with a single ladder, we can see from Fig. 10 that the double ladder possesses a lower threshold.

### B. Transconductance parametrization

As previously discussed, a good estimate for the $g_m$ needed to cancel losses and start oscillation is to measure the input admittance at a node where a negative conductance is to be inserted. From Fig. 6, $Y_{in}$ of the 5th cell is approximately 1.2 mS. Fig. 11 shows the minimum magnitude of the transconductance versus the number of cells of the double ladder.

The lossless double ladder's threshold follows the same trend as its quality factor varying as a function of length; i.e., since $Q_{tot} \propto N^5$ the double ladder's threshold $g_{m,\min} \propto 1/N^5$ when the active device is attached to every cell. In addition, it can be observed that when the active device is connected only to the middle cell, the threshold is higher and its trend versus $N$ is fitted to $g_{m,\min} \propto 1/N^{3.75}$. This is a remarkable feature of DBE resonators in general as compared to single-ladder oscillators in which scaling is $g_{m,\min} \propto 1/N^3$ when active devices are connected to *each* cell [16]. When element losses are considered here with $Q_e = 400$, the minimum $g_m$ increases from the lossless case, particularly for large $N$. However, we see that for large $N$ the threshold saturates, which is a consequence of the saturation feature of $Q_{tot}$ factor versus $Q_e$ seen previously in Fig. 5(a).

To provide a meaningful comparison we compare the threshold $g_{m,\min}$ at the $(l,n)=(2, N/2)$ node only, varying as a function of the $Q_e$ as well as $Q_{tot}$. For all of the remaining simulations, the driving point for negative resistance is set to be the $(N/2+1)^{th}$ cell. It is shown in Fig. 12 that for a given $Q_{tot}$, the double ladder exhibits a lower oscillation threshold than the single ladder.

Moreover, when comparing the threshold versus the element $Q_e$, the double ladder shows a lower threshold even for a relatively low element $Q_e <100$. The reason for the better behavior of the double ladder is that it features a fourth-order degeneracy. This implies that the circuit characteristics are more sensitive to perturbation in loss or gain, as in this case, compared to standard circuits or to a single ladder with a second-order degeneracy like the RBE. Fig. 12 shows how the threshold transconductance varies as a function of $Q_e$ for a double ladder and single ladder of two different sizes $N = 8$ and $N = 16$. We can see that for a low element quality factor, the 8-cell double-ladder circuit has the lowest threshold, for $Q_e > 100$. Furthermore, the single ladder of 16 cells has a higher threshold than all other configurations, which indicates indeed that a double ladder has an improved characteristic compared to a single ladder. This is a novel phenomenon that could be further investigated in order to enhance the efficiency of microwave oscillators.

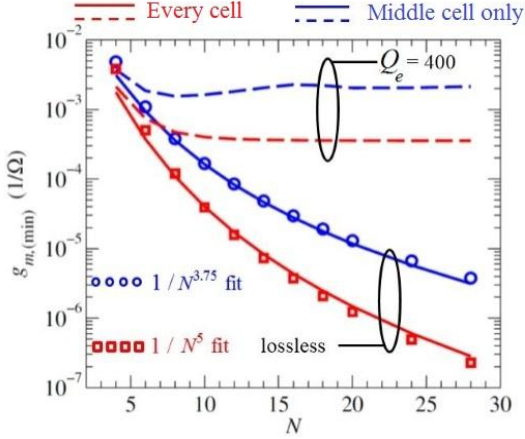

Fig. 11. Minimum $g_m$ (to start oscillations) scaling versus number of unit cells $N$ for the double ladder oscillator. The plot also shows for the lossless case a trend corresponding to $g_{m,\min} \propto 1/N^5$.

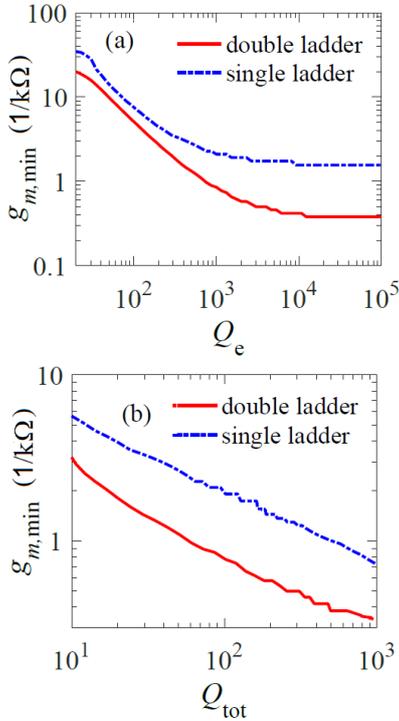

Fig. 12. Minimum $g_m$ (to start oscillations) versus different values for (a) quality factor $Q_e$ of elements, and versus (b) total $Q_{\text{tot}}$, for double and single ladder oscillators.

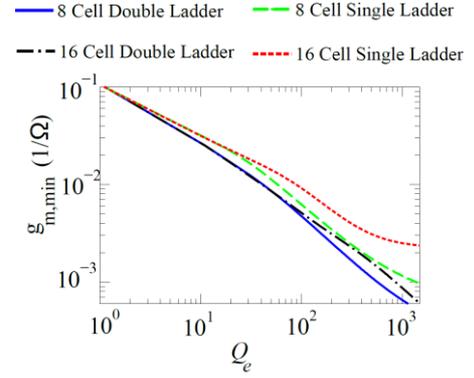

Fig. 13. Minimum $g_m$ varying as a function of the element quality factor for double ladders and single ladders of two different sizes $N = 8$ and $N = 16$.

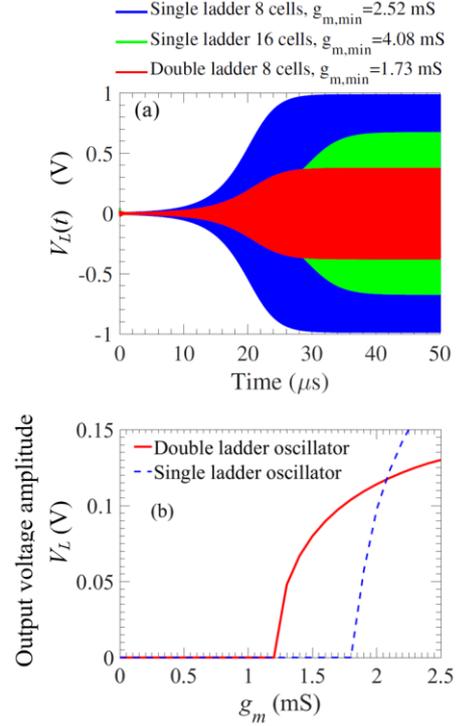

Fig. 14. (a) Transient load voltage for three different cases of single-ladder and double-ladder oscillators in comparison. (b) Steady-state output voltage amplitude for different values of $g_m$ comparing single and double ladders with $N = 8$ unit cells.

## V. THE DOUBLE LADDER OSCILLATOR

Here we study the time-domain response of the proposed double ladder oscillator (DLO). The transient behavior of this oscillator is simulated using Keysight ADS. The I-V characteristics of the active device was modeled as in (3) with $V_b = 1$ V. Fig. 14(a) shows one of the main advantages of the double ladder circuit over a single ladder: An 8-cell double ladder circuit requires 30% less $g_m$ than an 8-cell single ladder and 57% less $g_m$ than a 16-cell single-ladder for the circuit to oscillate. In addition, Fig. 14(b) shows the steady-state output voltage amplitude for single and double ladder oscillators (both made of 8 unit cells) versus $g_m$. The double-ladder oscillator produces higher load voltage amplitudes, in comparison to the single ladder counterpart, for small $g_m$ (above threshold) up to $g_m = 2.02$ mS, thus again showing potential advantages of low threshold oscillations, specifically in applications requiring low power

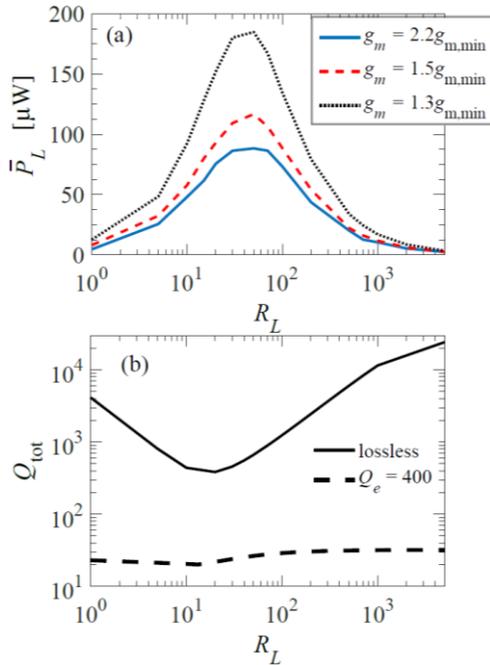

Fig. 15. (a) Average steady state power delivered to the load versus loading $R_L$, for $Q_e = 400$. (b) Minimum loaded quality factor corresponding to the lossless and lossy double ladder resonator.

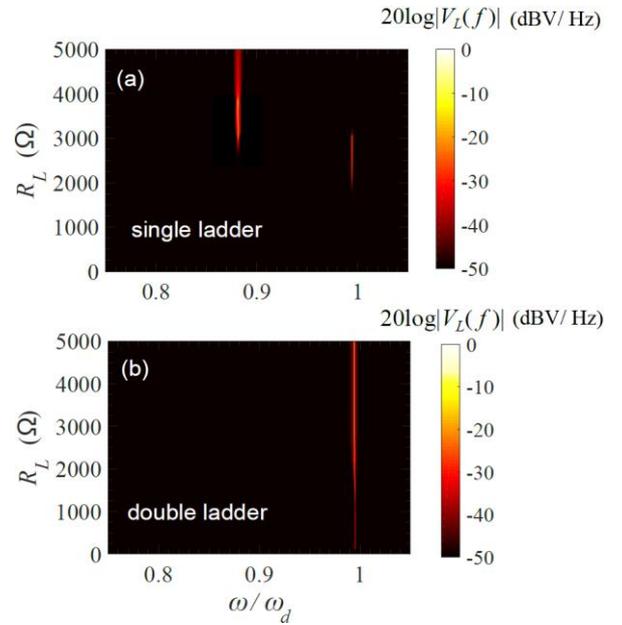

Fig. 16. Spectrum of load voltage $20\log|V_L(f)|$ for (a) double ladder oscillator and (b) single ladder oscillator varying as a function of the load resistance $R_L$, for $g_m$= 3 mS, with $Q_e = 400$.

consumption.

The average steady-state power delivered to the load $\bar{P}_L$ of the double-ladder circuit is depicted in Fig. 15 for three different values of $g_m$ as a function of the load resistance. We observe that the power delivered to the load peaks at the minimum of the loaded quality factor of the lossless circuit. This happens at the characteristic impedance of the double ladder, which for this circuit is $\sqrt{L/C} \cong 28.3\,\Omega$.

Another advantage of the double-ladder oscillator is demonstrated in terms of the sensitivity of the oscillation frequency versus variation in the load. Typically, the oscillation amplitude decreases (and for some cases the oscillator may not even operate) when the output termination resistance is changed from its nominal value. For this reason, often the output buffer stages for LC oscillators are needed to stabilize it against those variations. Moreover, single-ladder oscillators have been shown to demonstrate *mode jumping* where the oscillation frequency and mode of operation changes for various loads (see for example the analysis in [6], and the explanation of this behavior in [13] near the band edge). To demonstrate this effect in ladder oscillators, the steady-state load voltage calculated from transient simulations varying as a function of load is considered. The spectrum of the load voltage is then calculated by applying a windowed Fourier transform (rectangular window with width of 1 μs) on the saturated voltage waveform.

The single-ladder load voltage spectrum versus the load resistance, for $g_m = 3$ mS and $Q_e = 400$ with $N = 8$, is shown in Fig. 16(a). From this spectrum it can be seen that the oscillation frequency jumps from the desired frequency near the RBE of the single ladder (~100 MHz) to a lower frequency (~87 MHz) for sufficiently high load impedance. Moreover, this circuit does not oscillate for impedances below 1000 Ω. Indeed, the single ladder threshold for 50 Ω calculated in Fig. 12 confirms that $g_m = 3$ mS is below the threshold for such a load. In contrast with the double-ladder oscillator with $N = 8$ and $Q_e = 400$ the oscillation frequency is independent of the variation for the load resistance as seen from Fig. 16(b). Such a remarkable feature of DBE-based oscillators indicates the stability against load pulling. Moreover, the single ladder exhibits multimode oscillation at particular load values (for instance at 3 kΩ the single ladder oscillator can oscillate at both 0.99 $\omega_d$ and 0.88$\omega_d$) while the double ladder oscillates at only a single frequency at ~0.995$\omega_d$ demonstrating the unique mode selection scheme of double ladders. Moreover, the double ladder oscillators also generate the same oscillation frequency for higher values of $g_m$. These properties can also be observed when the load has a reactive component.

## VI. CONCLUSION AND REMARKS

A novel oscillator design, based on a periodic double-ladder circuit, has been presented in which a degenerate band edge (DBE) condition occurs. The passive behavior of the double-ladder resonator has been analyzed by considering the quality factor scaling over size and effects of loss. It has been shown that the conditions for oscillation are relaxed for the double-ladder as compared to the single-ladder circuit, thus allowing lower power dissipation from the negative conductance circuit that is required. Moreover, the stability of the oscillation frequency in the presence of variation in the load impedance has been shown to not exhibit mode-jumping behavior that can be observed in single ladders. This advantage can result in less power consumption for the active elements.

Comparison between the proposed double-ladder oscillator and a conventional LC-tank oscillator circuit will be carried out in the future that would account for phase noise, power consumption and other practical aspects. Because lumped elements are generally lossy and occupy large areas, implementing the double ladder in microstrip or waveguides circuits at microwave frequencies poses a potential improvement over conventional microwave sources designs. Another aspect for oscillator design is phase noise, in which double ladder oscillator topologies may offer potential advantages compared to conventional LC tank. Accordingly, DBE-based oscillator design can also avoid the need for stages of power-hungry current-mode logic buffers to reach acceptable oscillation amplitude at the low impedance termination.

## REFERENCES

[1] B. Van der Pol, "The nonlinear theory of electric oscillations," *Radio Eng. Proc. Inst. Of*, vol. 22, no. 9, pp. 1051–1086, 1934.
[2] B. Razavi, *Design of integrated circuits for optical communications*. John Wiley & Sons, 2012.
[3] G. W. Pierce, "Piezoelectric crystal resonators and crystal oscillators applied to the precision calibration of wavemeters," in *Proceedings of the American Academy of Arts and Sciences*, 1923, vol. 59, pp. 81–106.
[4] *Oscillation generator*. Google Patents, 1927.
[5] R. E. Collin, *Foundations for microwave engineering*. John Wiley & Sons, 2007.
[6] T. Endo and S. Mori, "Mode analysis of a multimode ladder oscillator," *Circuits Syst. IEEE Trans. On*, vol. 23, no. 2, pp. 100–113, 1976.
[7] A. Hajimiri, S. Limotyrakis, and T. H. Lee, "Jitter and phase noise in ring oscillators," *Solid-State Circuits IEEE J. Of*, vol. 34, no. 6, pp. 790–804, 1999.
[8] H. Wu and A. Hajimiri, "Silicon-based distributed voltage-controlled oscillators," *Solid-State Circuits IEEE J. Of*, vol. 36, no. 3, pp. 493–502, 2001.
[9] B. Razavi and R. Behzad, *RF microelectronics*, vol. 2. Prentice Hall New Jersey, 1998.
[10] U. L. Rohde, A. K. Poddar, and G. Böck, *The design of modern microwave oscillators for wireless applications: theory and optimization*. John Wiley & Sons, 2005.
[11] H.-A. Tanaka, A. Hasegawa, H. Mizuno, and T. Endo, "Synchronizability of distributed clock oscillators," *IEEE Trans. Circuits Syst. Fundam. Theory Appl.*, vol. 49, no. 9, pp. 1271–1278, 2002.
[12] H.-C. Chang, X. Cao, U. K. Mishra, and R. A. York, "Phase noise in coupled oscillators: Theory and experiment," *Microw. Theory Tech. IEEE Trans. On*, vol. 45, no. 5, pp. 604–615, 1997.
[13] J. Sloan, M. Othman, and F. Capolino, "Theory of Double Ladder Lumped Circuits With Degenerate Band Edge," *(accepted under revision) arXiv reprint: arXiv*, vol. 1609.00044, Aug. 2016.
[14] G. L. Matthaei, L. Young, and E. M. Jones, "Design of microwave filters, impedance-matching networks, and coupling structures. Volume 2," DTIC Document, 1963.
[15] A. I. Zverev, *Handbook of filter synthesis*. Wiley, 1967.
[16] A. B. Williams and F. J. Taylor, *Electronic filter design handbook*. McGraw-Hill, Inc., New York, NY, 1995.
[17] D. M. Pozar, *Microwave engineering*. John Wiley & Sons, 2009.
[18] J. D. Joannopoulos, S. G. Johnson, J. N. Winn, and R. D. Meade, *Photonic crystals: molding the flow of light*. Princeton university press, Princeton, NJ, 2011.
[19] J. P. Dowling, M. Scalora, M. J. Bloemer, and C. M. Bowden, "The photonic band edge laser: A new approach to gain enhancement," *J. Appl. Phys.*, vol. 75, no. 4, pp. 1896–1899, Feb. 1994.
[20] A. Figotin and I. Vitebskiy, "Slow light in photonic crystals," *Waves Random Complex Media*, vol. 16, no. 3, pp. 293–382, 2006.
[21] M. Othman and F. Capolino, "Demonstration of a Degenerate Band Edge in Periodically-Loaded Circular Waveguides," *IEEE Microw. Wirel. Compon. Lett.*, vol. 25, no. 11, pp. 700–702, 2015.
[22] A. J. Bahr, "A coupled-monotron analysis of band-edge oscillations in high-power traveling-wave tubes," *Electron Devices IEEE Trans. On*, vol. 12, no. 10, pp. 547–556, 1965.
[23] A. P. Kuznetsov, S. P. Kuznetsov, A. G. Rozhnev, E. V. Blokhina, and L. V. Bulgakova, "Wave theory of a traveling-wave tube operated near the cutoff," *Radiophys. Quantum Electron.*, vol. 47, no. 5–6, pp. 356–373, 2004.
[24] H.-Y. Ryu, S.-H. Kwon, Y.-J. Lee, Y.-H. Lee, and J.-S. Kim, "Very-low-threshold photonic band-edge lasers from free-standing triangular photonic crystal slabs," *Appl. Phys. Lett.*, vol. 80, no. 19, pp. 3476–3478, May 2002.
[25] L. B. Felsen and W. K. Kahn, "Transfer characteristics of 2n-port networks," in *Proceedings of the symposium on millimeter waves*, 1959, pp. 477–512.
[26] A. Figotin and I. Vitebskiy, "Gigantic transmission band-edge resonance in periodic stacks of anisotropic layers," *Phys. Rev. E*, vol. 72, no. 3, p. 036619, Sep. 2005.
[27] J. L. Volakis and K. Sertel, "Narrowband and Wideband Metamaterial Antennas Based on Degenerate Band Edge and Magnetic Photonic Crystals," *Proc. IEEE*, vol. 99, no. 10, pp. 1732–1745, Oct. 2011.
[28] M. A. K. Othman, F. Yazdi, A. Figotin, and F. Capolino, "Giant gain enhancement in photonic crystals with a degenerate band edge," *Phys. Rev. B*, vol. 93, no. 2, p. 024301, Jan. 2016.
[29] N. Apaydin, L. Zhang, K. Sertel, and J. L. Volakis, "Experimental Validation of Frozen Modes Guided on Printed Coupled Transmission Lines," *IEEE Trans. Microw. Theory Tech.*, vol. 60, no. 6, pp. 1513–1519, Jun. 2012.
[30] M. A. Othman, M. Veysi, A. Figotin, and F. Capolino, "Low Starting Electron Beam Current in Degenerate Band Edge Oscillators," *IEEE Trans Plasma Sci*, vol. 44, no. 6, pp. 918–929, 2016.
[31] A. Figotin and I. Vitebskiy, "Frozen light in photonic crystals with degenerate band edge," *Phys. Rev. E*, vol. 74, no. 6, p. 066613, Dec. 2006.
[32] C. Löcker, K. Sertel, and J. L. Volakis, "Emulation of propagation in layered anisotropic media with equivalent coupled microstrip lines," *Microw. Wirel. Compon. Lett. IEEE*, vol. 16, no. 12, pp. 642–644, 2006.
[33] J. R. Burr, N. Gutman, C. Martijn de Sterke, I. Vitebskiy, and R. M. Reano, "Degenerate band edge resonances in coupled periodic silicon optical waveguides," *Opt. Express*, vol. 21, no. 7, pp. 8736–8745, Apr. 2013.
[34] M. A. K. Othman, X. Pan, Y. Atmatzakis, and C. G. Christodoulou, "Experimental Demonstration of Degenerate Band Edge in Metallic Periodically-Loaded Circular Waveguide," *arXiv preprint*, vol. 1611, Nov. 2016.
[35] B. Jung and R. Harjani, "ΣHigh-frequency LC VCO design using capacitive degeneration," *Solid-State Circuits IEEE J. Of*, vol. 39, no. 12, pp. 2359–2370, 2004.
[36] A. Ghadiri and K. Moez, "A dual-band CMOS VCO for automotive radar using a new negative resistance circuitry," in *Circuits and Systems (MWSCAS), 2010 53rd IEEE International Midwest Symposium on*, 2010, pp. 453–456.
[37] S. Skogestad and I. Postlethwaite, *Multivariable feedback control: analysis and design*, vol. 2. Wiley New York, 2007.